\documentstyle[prl,preprint,aps]{revtex}

\begin{document}
\draft

\title{Fresh look on triality}

\author{M. Faber}

\address{Institut f\"ur Kernphysik, Technische Universit\"at Wien,
A--1040 Vienna}

\author{O. Borisenko, S. Mashkevich and G. Zinovjev}

\address{Institute of Theortical Physics, Ukrainian Academy of Science,
Kiev 143}
\maketitle

\begin{abstract}
Investigating the  $Z_3$ symmetry in Quantum Chromodynamics (QCD) we show
that full QCD with a vacuum of vanishing baryonic number does not lead to
metastable phases. Rather in QCD with dynamical fermions, the degeneracy of
$Z_3$ phases manifests itself in observables without open triality.
\end{abstract}

\vskip 1truecm
\begin{center}
IK--TUW--Preprint 9308401\\
August 1993
\end{center}

\vskip 1truecm

\vskip 1truecm

The Lagrangian in pure gluonic Quantum Chromodynamics (QCD) has global
$Z_3$ symmetry. It is very well known that in the low temperature phase
$Z_3$ symmetry leads to confinement of color charges and in the high
temperature phase $Z_3$ symmetry is spontaneously broken, color charges are
screened by the gluon field. In QCD with dynamical quarks the fermionic
contribution to the Lagrangian breaks $Z_3$ symmetry explicitly.

Recently, in refs.  \cite{Dixit_Ogilivie,Kajantie} it has been concluded on
the one hand that this explicit $Z_3$ symmetry breaking leads to the
existence of metastable states at arbitrary high temperatures.  On the
other hand the authors of refs. \cite{Weiss,Chen} believe that $Z_N$ phases
cannot be prepared as real macroscopic systems. In the following we will
try to remove some confusion and show that QCD with dynamical fermions has
degenerate $Z_3$ phases and therefore also ordered-ordered phase
transitions like pure gluonic QCD. Furthermore, we emphasize that QCD is
able to describe $Z_3$ symmetric systems only, i.e. systems of zero
triality. In order to introduce our concept we study first the pure gluonic
system and then turn to QCD with dynamical fermions.

In lattice QCD the gluon field appears in the form of $SU(3)$ matrices
$U_{x,\mu}$ which are defined on links $(x,\mu)$ of a four-dimensional
euclidean lattice. In finite temperature lattice QCD the correlation
function $<L(\vec{r}_1) \cdots L^*(\vec{r}_N)>$ of several Polyakov loops

\begin{equation} \label{Polyakov_loop}
L(\vec{r}) = \frac{1}{3} \; \rm{Tr} \prod_{t=1}^{N_t} U_{(\vec{r},t),0}
\end{equation}

\noindent can be connected \cite{Mac_Lerran} with the free energy F of N
infinitely heavy quarks $q$ or antiquarks $\bar{q}$ at the corresponding
positions $\vec{r}_1,\cdots , \vec{r}_N$ at temperature $T$ relative to the
free energy of the vacuum

\begin{equation} \label{free_energy}
F(q(\vec{r}_1),\cdots ,\bar{q}(\vec{r}_N)) = -T \, \rm{ln}<L(\vec{r}_1)
\cdots L^*(\vec{r}_N)>.
\end{equation}

\noindent The thermodynamical average $< \cdots >$ is computed using the
partition function

\begin{equation} \label{path_integral}
Z = \int{} {\cal D} [U] \; e^{-S[U]}
\end{equation}

\noindent as a ''sum'' over all gauge field configurations {U}. In pure
gluonic QCD for most problems the appropriate choice for S is the Wilson
action

\parbox{11cm} {\begin{eqnarray*}
S_G[U] & = & \beta \sum_{x, \mu < \nu} \left(1 - \frac{1}{3} {\rm Re} \;
{\rm Tr} \; U_{x, \mu \nu} \right), \quad \beta = \frac {6}{g^2},\\
U_{x, \mu \nu} & = & U_{x, \mu} U_{x+\hat{\mu}, \nu} U^{\dag}_{x+\hat{\nu},
\mu} U^{\dag}_{x, \nu} \, ,
\end{eqnarray*}} \hfill
\parbox{1cm}{\begin{eqnarray} \end{eqnarray}}

\noindent where the plaquettes $U_{x, \mu \nu}$ are built of four links in
the $\mu \nu$-plane of a four dimensional euclidean lattice.

The action $S_G[U]$ is $Z_3$ symmetric. This means that multiplication of
all links in direction $\mu=0$ in the three dimensional x,y,z-torus with
fixed t, e.g. $t=0$ by a $Z_3$ element leaves the action invariant. A
single Polyakov loop $L(\vec{r})$ transformes under $Z_3$ non-trivially
(its triality ${\cal T}$ is one), and therefore  the distribution of $L$
values can be used as an indicator for spontaneous breaking of $Z_3$
symmetry.

Let us now discuss the results of lattice calculations and their common
interpretation. Monte-Carlo calculations in lattice QCD show usually a
characteristic behaviour of the spatial average $L$ of $L(\vec{r})$ during
a Monte-Carlo simulation. In the confinement regime (low $\beta$) $L$
scatters symmetrically around zero in the complex plane. In the deconfined
phase there appear three $Z_3$ symmetric maxima of the Polyakov loop
distribution in $0^{\circ}$- and $\pm 120 ^{\circ}$-directions. As the
number of tunneling events between the maxima decreases with increasing
$\beta$ one commonly expects $L$ to be frozen in the thermodynamical limit
in one of the $Z_3$ directions and thus to get spontaneous $Z_3$ symmetry
breaking. The appearance of the three peak structure of the L-distribution
in the deconfinement regime is considered as a demonstration that
spontaneous symmetry breaking on a finite lattice can never happen exactly.
In the thermodynamical limit one may obtain $\rm{arg}<L> = 1, \pm \frac{2
\pi}{3}$. Besides the unpleasant fact that for $\rm{arg}(<L>) = \pm \frac{2
\pi}{3}$ one should reconsider the definition (\ref{free_energy}) the above
interpretration of Monte-Carlo results leads in full QCD to doubtful
consequences. Therefore, we prefer another point of view:
As $\int{} {\cal D} [U] \; e^{-S_G[U]}$ is $Z_3$ invariant and the Polyakov
loop (\ref{Polyakov_loop}) is not, one has to get
\begin{equation} \label{vanishing_Polyakov}
<L> = 0
\end{equation}
in the confined and also in the deconfined phase. There is no contradiction
between this statement and the spontaneous breaking of $Z_3$ symmetry in
the deconfined phase. It rather shows that the expectation value $<L>$ is
not the appropriate observable for the demonstration of spontaneous
breaking of $Z_3$ symmetry. One should rather use $<L(0) L^*({\vec r})>$
correlations or the ($Z_3$ symmetric)  $L$ distributions. The latter is
equivalent to the determination of Polyakov loops  in $Z_3$ invariant
representations (octet, decuplet, antidecuplet, etc). After this rather
trivial observation that for a $Z_3$ invariant Lagrangian observables of
non vanishing triality have expectation value zero we will turn to the more
interesting case of full QCD.

In full QCD with dynamical fermions the action S contains a fermionic
contribution $S_F$

\parbox{11cm} {\begin{eqnarray*}
&& S = S_G + S_F, \\
&& S_F = \frac{n_F}{4} \sum_{x,x^{\prime}} \bar{\psi}_x M_{x,x^{\prime}}
\psi_{x^{\prime}}, M_{x,x^{\prime}} = D_{x,x^{\prime}} + m
\delta_{x,x^{\prime}}
\end{eqnarray*}} \hfill
\parbox{1cm}{\begin{eqnarray} \end{eqnarray}}

\noindent which breaks $Z_3$ symmetry explicitly. In the Kogut-Susskind
formulation \cite{staggered_fermions} the fermionic matix M reads

\parbox{11cm} {\begin{eqnarray*}
&& M_{x,x^{\prime}} = \frac{1}{2} \sum_{\mu} \left( \Gamma_{x, \mu} U_{x,
\mu} \delta_{x^{\prime}, x+\mu} - \Gamma_{x^{\prime}, \mu}
U^{\dag}_{x^{\prime}, \mu} \delta_{x^{\prime}, x-\mu} \right) + m
\delta_{x,x^{\prime}}, \\
&& \Gamma_{x, \mu} = (-1)^{x_1+x_2+ \cdots +x_{\mu-1}},
\end{eqnarray*}} \hfill
\parbox{1cm}{\begin{eqnarray} \end{eqnarray}}

\noindent with one-component Grassmann variables $\psi_x$ and
$\bar{\psi}_x$. The integration over the Grassmann variables in the path
integral

\begin{equation}
Z = \int{} {\cal D} [U,\bar{\psi},\psi] \; e^{-S}
\end{equation}

\noindent can be performed analytically and leads to the fermionic
determinant ${\rm det} M$

\begin{equation} \label{fermionic_path_integral}
Z = \int{} {\cal D} [U] \; e^{-S_G[U]} {\rm det} M.
\end{equation}

It is easy to see that the fermionic action violates $Z_3$ symmetry
\cite{Banks}. In the usual MC-iterations one finds that in the low
temperature phase the expectation value $<L>$ of a Polyakov loop $L$ is a
small positive number and in the high temperature phase the three peak
structure of the distribution becomes asymmetric. The maximum at $0^\circ$
is favoured and the two lower maxima at angles of around $\pm120^\circ$ are
symmetric. This fact is usually interpreted as a consequence of the
violation of $Z_3$ symmetry.

Effective models based on this $Z_3$ asymmetric $L$-distribution are
discussed in refs. \cite{Dixit_Ogilivie,Kajantie,Weiss,Chen}. As mentioned
in the introduction they predict metastable states, but with unphysical
properties.

In order to resolve this contradiction we start from the statement that the
baryonic number $B$ of the QCD vacuum should be zero. It follows that also
the triality $\cal{T}$ is zero. Instead of the canonical ensemble with
$B={\cal T}=0$ which should be well suited, a grand canonical description
is allowed if it predicts the same behaviour in the thermodynamical limit.
Common lattice Monte-Carlo calculations in full QCD use the full fermionic
determinant ${\rm det} M$ and thus simulate a grand canonical ensemble with
chemical potential $\mu=0$ containing all  three triality sectors. There
appears now the following loophole in the usual argumentation. The
determination of the Polyakov loop expectation value in full QCD gives $<L>
\ne 0$ and a $Z_3$ asymmetric $L$-distribution. In assuming to have
measured the $L$-distribution of the vacuum several authors use effective
potentials for the time component of the gluon field arising from one loop
effects and predict metastable states \cite{Dixit_Ogilivie,Kajantie} or
find properties of these states \cite{Weiss,Chen} which are not possible in
Minkowski space. However, it has been overlooked that the determinant ${\rm
det} M$ may be decomposed in three different triality sectors. The
decomposition can easily be derived in the following way: Via a hopping
parameter expansion the fermionic determinant can be decomposed into
contributions according to the number of closed loops of gauge field U
which wind around the 4-dimensional lattice in time
direction\cite{Weiss2,Barbour,Hasenfratz}. We can then define the triality
operator $\hat{\cal{T}}$ (with eigenvalues $0, \pm 1$) which determines the
triality of such closed loops, and with his help a projection operator
$\hat{P}_{\cal T}$ on triality $\cal T$
\begin{equation}
\hat{P}_{\cal T} = \frac{1}{3} \sum_{q=0,\pm 1} e^{q 2 \pi i (\hat{\cal T}
- {\cal T})/3}.
\end{equation}
Then the triality zero contribution of the fermionic determinant is
\begin{equation}
{\rm det}_0 M = \hat{P}_0 {\rm det} M = \frac{1}{3} \sum_{q=0,\pm 1} e^{q 2
\pi i \hat{\cal T}/3} {\rm det} M
\end{equation}
\renewcommand{\thefootnote}{\fnsymbol{footnote}}
and more general a projection operator on triality $\cal T$
\footnote[4]{The implementation of this projection operator in the lattice
formulation is, at least in principle, very simple. One has to multiply all
links in time direction for an arbitrary time step with the phase $e^{q 2
\pi i /3}$ and to determine with this new link variables the determinant.
The result has to be multiplied with $e^{q 2 \pi i {\cal T}/3}$. Finally,
the sum over  phases $q = 0, \pm 1$ has to be executed.}
\renewcommand{\thefootnote}{\arabic{footnote}}
\begin{equation}
\hat{P}_{\cal T} = \frac{1}{3} \sum_{q=0,\pm 1} e^{q 2 \pi i (\hat{\cal T}
- {\cal T})/3}.
\end{equation}
and in general
\begin{equation} \label{detM}
{\rm det} M = \sum_{{\cal T}=0,\pm 1} {\rm det}_{\cal T} M,
\end{equation}
with
\begin{equation}
{\rm det}_{\cal T} M = \hat{P}_{\cal T} {\rm det} M.
\end{equation}

{}From eq. (\ref{detM}) it is obvious that the observable $L$ having triality
${\cal T}=1$ tests only the ${\cal T}=-1$ sector of the fermionic
determinant which is not the vacuum sector. Using the correct ${\cal T}=0$
vacuum sector for the fermionic determinant results in a $Z_3$ symmetric
$L$-distribution, an expectation value $<L>=0$ and no prediction of
metastable and unphysical vacuum states. Let us discuss this in more
detail. As the measure ${\cal D} [U]$ and the gluonic Lagrangian $S_G[U]$
are $Z_3$ symmetric, the product ${\rm det} M[U] O_{\cal T}[U]$ in
\begin{equation}
<O_{\cal T}> = \frac{1}{Z} \int {\cal D} [U] e^{-S_G[U]} {\rm det} M[U]
O_{\cal T}[U]
\end{equation}
has to be also $Z_3$ symmetric to get $<O_{\cal T}> \ne 0$. In other words
$Z_3$ violating contributions in ${\rm det} M[U] O_{\cal{T}}[U]$ are
automatically eliminated by the path integral. Only the triality $-T$
component ${\rm det}_{-{\cal T}} M$ of the fermionic determinant survives.
${\rm det}_{-{\cal T}} M$ has $T$ quark loops less than antiquark loops
winding around the lattice in time direction.

Applying this statement to the example of the single Polyakov loop $L$
representing an infinitely heavy quark $Q$, the fermionic determinant
supplies a triality $-1$ state of light quarks which will mostly consist of
a light antiquark $\bar{q}$ in order to color neutralize the heavy quark Q.
A finite value of $<L>$ in full QCD therefore does not mean that $Z_3$
symmetry is broken but rather that the fermionic system is flexible enough
to neutralize external charges. The Polyakov expectation value $<L>$ is a
thermodynamical mixture
\begin{equation}
<L> = ... + e^{-F(Q\bar{q})/T} + e^{-F(Qqq)/T} + ...
\end{equation}
of a heavy-light meson $Q\bar{q}$, a heavy-light baryon $Qqq$, ... . This
heavy-light states should be  bound in the hadronic and "ionized" in the
quark-gluon plasma phase. The charge density of such light quarks around
static sources has been measured in ref. \cite{Mueller}.

For the system of a heavy quark $Q$ at position $\vec{r}_1$ and a heavy
antiquark $\bar{Q}$ at position $\vec{r}_2$
\begin{equation}
L(\vec{r}_1) L^*(\vec{r}_2)
\end{equation}
a totaly different component of the fermionic determinant, the triality
zero component, contributes. It is evident that this component is $Z_3$
symmetric, i.e. the three $Z_3$ transformed sectors which differ in $L$ by
factors $e^{\pm \frac{ 2 \pi i}{3}}$ contribute with the same Boltzmann
factor. This example shows that for observables of triality zero full QCD
is $Z_3$ symmetric. The fermionic action does not destroy $Z_3$ symmetry as
is the general belief.

We can even extend this statement for triality non zero observables. The
above discussion for a single Polyakov line has demonstrated that for $Z_3$
violating observables full QCD restores the $Z_3$ symmetry by compensating
them to triality zero. Therefore, the effective models of refs.
\cite{Dixit_Ogilivie,Kajantie,Weiss,Chen} are effective models of
heavy-light hadrons in the QCD vacuum and not effective models of "the" QCD
vacuum.

We would like to summarize that the vacuum of QCD with dynamical fermions
has triality zero and therefore degenerate $Z_3$ phases and ordered-ordered
phase transitions like pure gluonic QCD. The Polyakov loop can still be
used as an order parameter for the breaking of $Z_3$ symmetry. There is no
need to search new order parameters \cite{DeTar} for full unquenched QCD.
In usual lattice Monte-Carlo simulations this triality zero sector can be
projected out by using triality zero obsevables only. Another method would
be to use ${\rm det}_0 M$ as the fermionic Boltzmann factor. However, an
implementation of a projection to fixed triality $\cal T$ (e.g. ${\cal T} =
0$) will be rather complicated due to the fact that usually not ${\rm det}
M$ is simulated but $({\rm det} M)^2$ whose ${\cal T} = 0$ component
\begin{equation}
({\rm det} M)^2_{{\cal T} = 0} = ({\rm det}_0 M)^2 + 2 {\rm det}_1 M   {\rm
det}_{-1} M
\end{equation}
contains contributions from triality $\pm 1$ components. Therefore, one
will have to rely on common lattice calculations with the limitations that
one should not be mislead by the mixing of different triality sectors in
the grand canonical ensemble.

M.F. thanks for the warm hospitality at the Institute of Theoretical
Physics of the Ukrainian Academy of Sciene, where the idea of this work
came up and to \v S.~Olejn\'\i k for critical reading of the manuscript.
G.M.Z. wishes to thank J.~Engels, B.~Petersson and H.~Satz for interesting
discussions. The work was partially supported by
''Fonds zur F\"orderung der wissenschaftlichen Forschung'' under
Contract~No.~P7237-TEC.

After finishing the first version of this article we noticed that J.
Polonyi \cite{Polonyi} introduced a concept similar to ours. The main
conceptual difference to Polonyi's idea is that we demonstrate that common
Monte-Carlo calculations use the correct Boltzmann factor for the fermionic
vacuum if only triality zero observables are measured.


\begin{references}

\bibitem{Dixit_Ogilivie} V.~Dixit and M.~C.~Ogilvie, Phys. Lett. B {\bf
269}, 353 (1991).
\bibitem{Kajantie} J.~Ignatius, K.~Kajantie and K.~Rummukainen, Phys. Rev.
Lett. {\bf 68}, 737 (1992).
\bibitem{Weiss} V.~M.~Belyaev, Ian~I.~Kogan, G.~W.~Semenoff and N.~Weiss,
Phys. Lett. B {\bf 277}, 331 (1992).
\bibitem{Chen} W.~Chen, M.~I.~Dobroliubov and G.~B.~Semenoff, Phys. Rev. D
{\bf 46}, R1223 (1992).
\bibitem{Mac_Lerran} L.~D.~McLerran and B.~Svetitsky, Phys. Rev. D {\bf
24}, 450 (1981).
\bibitem{staggered_fermions} J.~B.~Kogut, Rev. Mod. Phys. {\bf 55}, 755
(1983).
\bibitem{Banks} T.~Banks and A.~Ukawa, Nucl. Phys. B {\bf 225}, 145 (1983).
\bibitem{Weiss2} N.~Weiss, Phys. Rev. D {\bf 35}, 2495 (1987).
\bibitem{Barbour} I.~B.~Barbour, C.~T.~H. Davies and Z.~Sabeur, Phys. Lett.
B {\bf 215}, 567 (1988).
\bibitem{Hasenfratz} A.~Hasenfratz and D.~Touissant, Nucl. Phys. B {\bf
371}, 539 (1992).
\bibitem{Mueller} W.~B\"urger, M.~Faber, H.~Markum and M.~M\"uller, Phys.
Rev. D 47, 3034 (1993).
\bibitem{DeTar} C.~DeTar and L.~D.~Mc Lerran, Phys. Lett. B {\bf 119}, 171
(1982).
\bibitem{Polonyi} J.~Polonyi, in {\it Selected Topics in Quark
Confinement}, edited by J.~Polonyi, F.~Csikor, A.~Patk\'os and
K.~Szlach\'anyi (E\"otv\"os University, Budapest, 1992).
\end{references}
\end{document}